

\magnification=\magstep1

\font\bigbf=cmbx10 at 12pt
\def\case#1/#2{{\textstyle {#1\over#2}}}
\def\cite#1{[#1]}
\count20=0
\count21=0
\def\section#1{\medskip
\advance\count21 by 1
{\bf\noindent\number\count21. #1}
\smallskip
\count20=0}
\def\begin{\advance\count20 by 1 $$}
\def\endeq{\eqno{(\number\count21.\number\count20)}$$}
\def\ref#1{(#1)}

\hfill{NCL93--TP10}
\bigskip
\centerline{\bigbf Stability of the Cauchy horizon}
\centerline{\bigbf in Kerr--de Sitter spacetimes}
\bigskip \centerline{\bf Chris M. Chambers and Ian G. Moss}
\medskip
\centerline{\bf Department of Physics}
\centerline{\bf University of
Newcastle upon Tyne}
\centerline{\bf Newcastle upon Tyne, NE1 7RU}
\centerline{\bf U.K.}
\bigskip \bigskip
\centerline{ABSTRACT}
\medskip
{\narrower
We begin a program of work aimed at examining the interior of a
rotating black hole with a non--zero cosmological constant. The
generalisation of Teukolsky's equation for the radial mode functions is
presented. It is shown that the energy fluxes of scalar,
electromagnetic and gravity waves are regular at the Cauchy horizon
whenever the surface gravity there is less than the surface gravity at
the cosmological horizon. This condition is narrowly allowed, even when
the cosmological constant is very small, thus permitting an observer to
pass through the hole, viewing the naked singularity along the way.}
\vfill
\eject
\section{Introduction}

The fate of an observer who falls into a black hole is an interesting
and important issue within the context of general relativity. The
situation of interest is when the black hole is charged or rotating,
because then the spacetime can be continued beyond the black hole into
another universe, with naked singularities and causality violation to
contend with.

We know that for a black hole in empty space the observer's progress is
impeded by a singularity at the Cauchy horizon \cite{1,2}, but the
situation is different if there is a cosmological constant. In the
non--rotating case we now know that there is a finite range of
parameters for which the Cauchy horizon is fully stable \cite{3--6}. In
these cases an observer could pass into the hole and see the naked
singularity. Since a cosmological constant does not contradict the laws
of physics in any known way, this journey constitutes at the very least
a valid thought experiment in which the predictive power of general
relativity breaks down. The evolution of the spacetime beyond the
Cauchy horizon requires a knowledge of `initial' conditions at the
singularity and this lies beyond the limitations of current theory.

In this paper we examine the stability of the Cauchy horizon for a
rotating hole when a cosmological constant is present by consideration
of the energy fluxes of scalar, electromagnetic and gravity waves
propagating on the black hole background. We begin by separating the
field equations into equations for radial and angular functions. For
this step in the calculation the introduction of a cosmological
constant becomes an asset because it makes manifest the symmetry of the
metric under the interchange of the radial coordinate $r$ and the
coordinate $\mu=a\cos\theta$. It is the strength of this symmetry
combined with the algebraic character of the curvature tensor that
makes possible the separation of the fields and ultimately makes the
problem solvable.

The significance of gravitational wave equations to the stability lies
in the fact that they result from perturbing the full set of Einstein's
field equations \cite{1}. Therefore an arbitrarily small perturbation,
provided it does not grow in size anywhere in the spacetime region of
interest, can be analysed by purely linear equations. This is to be
contrasted with the unstable case, familiar from the ordinary Kerr metric,
where it is necessary to solve non--linear equations because an initially
small perturbation becomes large near to the Cauchy horizon \cite{2}.

As a matter of fact, gravitational wave propagation is only part of
the full gravitation perturbation analysis because there are other
degrees of freedom in the metric, but it ought to contain all of the
physical degrees of freedom relevant to the stability question.
To be certain, it only remains to be shown that the remaining
gravitational perturbation equations separate into similar wave
equations to deduce stability in the full gravity--matter system.
We hope to establish the complete separation of the equations in a
later paper.

We find stability requires that the surface gravity $\kappa$ of the
cosmological horizon should be larger than the surface gravity of the
Cauchy horizon; the same condition as in the charged case. The
parameter range for which this condition holds are found in the
following section, in which various properties of the spacetime are
explained. The parameter range is small, but finite.

\section{The spacetime}

The metric of a stationary black hole in de Sitter space was first
given by Carter \cite{7} and takes the form,
\begin
ds^2=\rho^2(\Delta_r^{-1}dr^2+\Delta_\mu^{-1}d\mu^2)
+\rho^{-2}\Delta_\mu\,\omega^1\otimes\omega^1
-\rho^{-2}\Delta_r\,\omega^2\otimes\omega^2,
\endeq
with one--forms,
\begin
\chi^2\omega^1=dt-a^{-1}\sigma_r^2d\phi\hbox{,~~~~}
\chi^2\omega^2=dt-a^{-1}\sigma_\mu^2 d\phi.
\endeq
The metric is parameterised as follows,
\begin\eqalign{
&\sigma_r=(a^2+r^2)^{1/2}\hbox{,~~~~~~}
\Delta_r=(a^2+r^2)(1-\case1/3\Lambda r^2)-2Mr+Q^2\cr
&\sigma_\mu=(a^2-\mu^2)^{1/2}\hbox{,~~~~}
\Delta_\mu=(a^2-\mu^2)(1+\case1/3\Lambda \mu^2)
}
\endeq
and
\begin
\rho^2(r)=r^2+\mu^2\hbox{,~~~~}
\chi^2=1+\case1/3\Lambda a^2\hbox{,~~~~}
\Omega(r)=a/(r^2+a^2).
\endeq
The function $\Omega$ is explained below.
Note that the subscripts are used to denote functions that depend
only on the coordinate indicated by the subscript, and a prime will
always denote differentiation with respect to this coordinate.

The metric is a solution of the Einstein--Maxwell equations with a
non--zero cosmological constant $\Lambda$ and the electromagnetic
vector potential of a charge $Q$,
\begin
A={Q\,r\over \rho^2}\omega^2.
\endeq
To see that this metric really does represent a black hole we introduce
the angle $\theta$, where
$\mu=a\cos\theta$. The surfaces of constant radius $r$ and time $t$ are
distorted spheres, parameterised by spherical polar coordinates
$\theta$ and $\phi$ in the usual way.

The metric has coordinate singularities at the roots of $\Delta_r=0$. For
a range of parameters shown in figure 1 there are four real roots,
which shall be labelled in decreasing order by $r_1$, $r_2$, $r_3$ and
$r_4$. The roots $r_2$ and $r_3$ represent the outer and inner horizons
respectively of the black hole, whilst $r_1$ represents a de Sitter
horizon. The fourth root $r_4$ is negative and non--physical.

It is possible to set up coordinate systems for which the metric is
regular on any particular horizon. Following section 56 in ref. \cite{1},
in the region outside the event horizon there exists two simple families of
null geodesics with tangent vectors
\begin
\dot t=\chi^4{\sigma_r^2\over\Delta_r}\omega,\quad
\dot r=\pm\chi^2\omega,\quad
\dot\mu=0\hbox{,~~~~and~}
\dot\phi=\chi^4{a\over\Delta_r}\omega,
\endeq
where a dot denotes a derivative with respect to the geodesic
affine parameter and $\omega$ is a constant. Corresponding `light--like'
coordinates which are fixed along each family can therefore be defined
by $u=t-r^*$ and $v=t+r^*$, where
\begin
dr^*=\chi^2\Delta_r^{-1}\sigma_r^2dr.
\endeq

Additionally, for rotating black holes, there is a unique combination
of the killing vectors that is null on the (future) event horizon,
$l=\partial_t+\Omega(r_2)\partial_\phi$. This leads us to introduce
an azimuthal angle coordinate
\begin
\phi_2=\phi-\Omega(r_2)t
\endeq
on the horizon, which is constant along the integral curves of $l$.
Physically, this corresponds to the dragging of inertial frames near
to the hole.

The null--killing vector on the horizon $r_i$ can be used also to
define the surface gravity, $\nabla\,l^2=\kappa_il$. The explicit
expression for $\kappa_i$ follows from squaring both sides of the
expression and taking the limit as $r\to r_i$ \cite{8},
\begin
\kappa_i=\case1/6\Lambda\chi^{-2}\sigma_r^{-2}
\prod_{i\ne j}|r_i-r_j|.
\endeq
(There is a typographical mistake in ref. \cite{9}).

The extension of the metric through the event horizon was given by
Carter in ref \cite{7} (with some corrections given in ref. \cite{8}).
The regular coordinate system is given in terms of the surface gravity,
\begin
U=-\kappa_2^{-1}e^{-\kappa_2 u}\hbox{,~~~~}
V=\kappa_2^{-1}e^{\kappa_2v}
\endeq
The coordinates $u$ and $v$ will be defined inside the event horizon as
shown in figure 2.

It is possible to extend the metric through each of the horizons in
turn and obtain the complete Penrose diagram, a part of which is shown
in figure 3 \cite{7,8}. The fully extended spacetime
stretches arbitrarily far in each direction. It is possible to find
world--lines that pass through the black hole from region $I$, through
regions $II$ and $III$ and on to the outside.

The exterior geometry of a collapsing star would only take in part
of the picture, the remainder being replaced by the geometry of the
stellar interior. In order to retain the axial symmetry of the solution
it is also necessary to place a second body of equal mass around the
point of the universe that is antipodal to the centre of the star.
This second body can be thought of as representing the rest of the
matter in the universe.

Inside region $III$ lies the naked singularity and the possibility
of causality violation from closed timelike curves. These curves pass
beyond the singularity at $r=0$, which only appears in the $\mu=0$ plane,
into the next region in the Penrose diagram. The problem for an
observer who tries to enter these regions lies in crossing
the Cauchy horizon between regions $II$ and $III$ at $r=r_3$.

As an observer crosses the Cauchy horizon any radiation from outside
appears to be either red--shifted or blue--shifted depending on the
relative values of the surface gravity at the Cauchy and de Sitter
horizons, red--shifted if the Cauchy horizon value is the smaller. In
the non--rotating case, this was also the Cauchy horizon stability
condition \cite{4,5}.

It is possible to split up the parameter space of the black hole
metrics into regions depending on the relative sizes of the surface
gravity on various horizons. We write the difference between the
surface gravities at $r_i$ and $r_j$ as
\begin
\kappa_i-\kappa_j=\case1/6\Lambda\chi^{-2}
{|r_i-r_j|\over(r_i^2+a^2)(r_j^2+a^2)}p_{ij},
\endeq
where $p_{ij}$ is a polynomial in the radii.

If two of the radii are equal, then the surface gravities of the
respective horizons both vanish. In other cases the relative magnitudes
of the surface gravities are determined by the signs of the
polynomials. Explicit expressions for these polynomials are,
\begin
\eqalign{
p_{12}&=(r_1r_2+r_3r_4-2a^2)(r_1^2-r_2^2)\cr
p_{13}&=(r_1r_3+r_2r_4)(r_1+r_3)^2+2(r_1r_3-a^2)(r_1r_3-r_2r_4).}
\endeq

The black--hole parameters for which $p_{ij}=0$ can be found by first
writing  $p_{ij}$ in terms of $x=r_1r_2-r_3r_4$ or $x=r_1r_3-r_2r_4$ in
the case of $p_{12}$ or $p_{13}$ respectively. From the conditions on
roots of $\Delta_r=0$ it is possible to deduce quite independently that
$x$ satisfies
\begin
M^2=\case1/{36}\Lambda^2x^3+\case1/{12}\Lambda(1-\case1/3\Lambda
a^2)x^2
+\case1/3\Lambda(a^2+Q^2)x
+(1-\case1/3\Lambda a^2)(a^2+Q^2).
\endeq
Solving $p_{ij}=0$ for $x$ gives an expression $x(a,Q,\Lambda)$ which
can be substituted in this equation to give a condition
$M(a,Q,\Lambda)$ for equal surface gravities.

The situation in which the surface gravities on the de Sitter and event
horizons are equal gives rise to gravitational instantons and was
discussed in reference \cite{8}. From equation \ref{2.12}, the solution
for this case is $x=2a^2$ and leads to a condition from \ref{2.13},
\begin
M^2=Q^2\chi^2+a^2\chi^4.
\endeq

The borderline of the stability criterion is that $\kappa_1=\kappa_3$.
This can be found by solving $p_{13}=0$ numerically and substituting
the result into \ref{2.13}, leading to the
condition shown in figure 1. The line lies very close to the line where
the inner and outer horizons coincide. The region of stability between
the two lines OA is small,
but finite in a uniform measure on the space of parameters $M$, $Q$ and
$a$ for fixed $\Lambda$.

\section{The scalar wave equation}

We will take a massless scalar field $\Phi$, with wave equation
\begin
\nabla^2\Phi=0,
\endeq
to demonstrate how the wave--equation separates on the black hole
background \cite{10}.

In terms of coordinates the Laplacian becomes
\begin
\nabla^2={1\over\rho^2}\partial_r\Delta_r\partial_r
+{1\over\rho^2}\partial_\mu
\Delta_\mu\partial_\mu
+{\rho^2\over\Delta_\mu}\partial_1^2
-{\rho^2\over\Delta_r}\partial_2^2
\endeq
where $\partial_1$ and $\partial_2$ are dual vectors to $\omega_1$ and
$\omega_2$,
\begin
\partial_1=-{\chi^2\over\rho^2}(\sigma_\mu^2\,\partial_t+a\partial_\phi)
\hbox{~~~~~}
\partial_2={\chi^2\over\rho^2}(\sigma_r^2\,\partial_t+a\partial_\phi).
\endeq

We look for separable solutions of form
\begin
\Phi=R(r)S(\mu)e^{-i\omega t}e^{im\phi}.
\endeq
The angular and radial terms in the wave equation separate immediately,
with  angular part
\begin
-\partial_\mu\Delta_\mu\partial_\mu S
+{K_\mu^2\over\Delta_\mu}S=\lambda S,
\endeq
where
\begin
K_\mu=\chi^2(am-\sigma_\mu^2\omega).
\endeq

We require that the solutions be regular at $\mu=\pm a$. Scaling $\mu$
by $a$ shows that the eigenvalues $\lambda$ depend upon the parameters
$m$, $\omega a$ and $\Lambda a^2$. We also add another parameter $l$ to
label the position of the eigenvalue in sequence, starting from the
value $|m|$. The corresponding eigenfunctions will then be denoted by
$S(lm\omega;\mu)$. Some values of the angular eigenvalues obtained
numerically are tabulated in tables 1--3 for various parameter values.
The cosmological constant has been re--expressed there as
$\Lambda=3/\alpha^2$.

For $\Lambda=0$ the equation reduces to the equation for spheroidal
wave functions,
\begin
\left(\left(a^2-\mu^2\right)S'\right)'+
\left[\lambda+2am\omega-\omega^2(a^2-\mu^2)
-a^2m^2(a^2-\mu^2)^{-1}\right]S=0.
\endeq
The spheroidal eigenvalues denoted by $\lambda_l^m$ \cite{11} are then
related to $\lambda$ by a shift of $2am\omega$. We shall be interested
later in the analyticity properties of the angular functions in a strip
of the complex $\omega$ plane below the real axis. Numerical studies
for spheroidal wave functions indicate that there are branch cuts away
from the real axis and suggest that these functions are regular in the
region $|\omega|<c\,l^2$ for some constant $c$ \cite{11}.

For small values of $\Lambda$ it is possible to relate the properties
of the angular functions to the spheroidal wave functions by a suitable
approximation scheme. The angular functions are the stationary points
of an integral,
\begin
I[S]=\int_{-a}^ad\mu\left\{\Delta_\mu
(S')^2+(K_\mu^2/\Delta_\mu)S^2\right\}.
\endeq
We can separate the $\Lambda=0$ part and write
\begin
I=\sum_n I^{(n)}[S,S](\Lambda/3)^n,
\endeq
where
\begin
\eqalign{
I^{(0)}[S_1,S_2]&=\int_{-a}^ad\mu\left\{\sigma_\mu^2S_1'S_2'+
\sigma_\mu^{-2}(am-\sigma_\mu^2\omega)^2S_1S_2\right\}\cr
I^{(1)}[S_1,S_2]&=
\int_{-a}^ad\mu\left\{\mu^2\sigma_\mu^2S_1'S_2'+(1+a^2/\sigma_\mu^2)
(am-\sigma_\mu^2\omega)^2S_1S_2\right\}\cr
I^{(n)}[S_1,S_2]&=\int_{-a}^ad\mu\,(-\mu^2)^{n-1}\sigma_\mu^2
(am-\sigma_\mu^2\omega)^2S_1S_2.
}\endeq
These give series for small $\Lambda$,
\begin
\lambda_l=\sum_n \lambda_l^{(n)}(\Lambda/3)^n
\hbox{,~~~~}S_l=\sum_n S_l^{(n)}(\Lambda/3)^n.
\endeq
where $S_l^{(0)}$ is the spheroidal wave function. Thus
\begin
\lambda_l^{(1)}=I^{(1)}[S_l^{(0)},S_l^{(0)}]\hbox{,~~~~}
S_l^{(1)}=\sum_{q\ne l}(\lambda_l^{(0)}-\lambda_q^{(0)})^{-1}
I^{(1)}[S_l^{(0)},S_q^{(0)}]\,S_q^{(0)}.
\endeq
This result was used to check the values in the tables. Continuing
further with the series, it is possible to show that the $S_l^{(n)}$ are
regular functions of $\omega$ when $S_l^{(0)}$ is regular.

Returning now to the wave equation and taking the radial part,
\begin
-\partial_r\left(\Delta_r\,\partial_rR\right)
+\lambda R-{K_r^2\over\Delta_r}R=0,
\endeq
where
\begin
K_r=\chi^2(am-\sigma_r^2\omega).
\endeq

It is possible to write
\begin
\sigma_r\partial_r\left(\Delta_r\,\partial_r R\right)=
\left({\chi^4\sigma_r^4\over\Delta_r}{d^2\over d
r^{*2}}\right)\sigma_rR
-\left({r\Delta_r\over\sigma_r^3}\right)'\sigma_r^2R
\endeq
This allows the radial equation to be written in the form of a
scattering problem,
\begin
\left({d^2\over d r^{*2}}+V(r)\right)\sigma_rR=0
\endeq
where the potential $V(r)$ takes the form
\begin
V=\left(\omega-m\Omega\right)^2
-{\lambda\Delta_r\over\chi^4\sigma_r^4}
-{\Delta_r\over\chi^4\sigma_r^3}\left({r\Delta_r\over\sigma_r^3}\right)'.
\endeq
This potential has been plotted as a function of $r^*$ between the
outer and inner horizons in figure 4. There is a potential well in the
region covered by the plot and a potential barrier outside of the event
horizon. The main difference from the non--rotating case is that the
values of the potential as $r^*\to\infty$ and $r^*\to-\infty$ need no
longer be equal.
\bigskip
\section{The electromagnetic wave equations}

In this section we will examine the Maxwell field equations on a
Kerr-de Sitter background using the Newman--Penrose \cite{12}
formalism. This development  follows the lines of the analysis for a
metric with vanishing cosmological constant \cite{13--15}. The main
difference lies in exploiting to the full the symmetrical appearance of
the metric with respect to $r$ and $\mu$.

We begin with an orthonormal tetrad of vector fields,
\begin
\eqalign{
&e_r={\sqrt{\Delta_r}\over\rho}\partial_r\hbox{,~~~}
e_2={\chi^2\over\rho\sqrt{\Delta_r}}
(\sigma_r^2\partial_t+a\,\partial_\phi)\cr
&e_\mu={\sqrt{\Delta_\mu}\over\rho}\partial_\mu\hbox{,~~~}
e_1=-{\chi^2\over\rho\sqrt{\Delta_\mu}}
(\sigma_\mu^2\partial_t+a\,\partial_\phi).
}
\endeq
The vectors dual to $\omega^1$ and $\omega^2$ have been used as in
section 2. Newman--Penrose formalism consists of writing the connection
forms in a complex null--tetrad basis,
\begin
\eqalign{
&l=\case1/{\sqrt{2}}(e_r+e_2)\hbox{,~~~~}
n=-\case1/{\sqrt{2}}(e_r-e_2)\cr
&m=\case1/{\sqrt{2}}(e_\mu+i\,e_1)\hbox{,~~~~}
\bar m=\case1/{\sqrt{2}}(e_\mu-i\,e_1).
}
\endeq
These vectors satisfy $l.n=-1$ and $m.\bar m=1$. (Inside the event
horizon we will use the modulus of $\Delta_r$ for scaling the tetrad
vectors and choose $l$ always to be in the direction of $\partial_V$,
where $V$ is the Kruskal coordinate.)

The notation for the covariant derivatives along the tetrad vectors
consists of
\begin
D=\nabla_l\hbox{,~~~}
\Delta=\nabla_n\hbox{,~~~}
\delta=\nabla_m\hbox{,~~~}
\bar\delta=\nabla_{\bar m}.
\endeq
The connection forms are then expressed as in the following table,
\begin
\eqalign{
&Dl=
(\epsilon+\bar\epsilon)l-\bar\kappa m-\kappa\bar m\hbox{,~~~}
Dm=
(\epsilon-\bar\epsilon)m-\kappa n+\bar\pi l,\cr
&\Delta l=
(\gamma+\bar\gamma)l-\bar\tau m-\tau\bar m\hbox{,~~~}
\Delta m=
(\gamma-\bar\gamma)m-\tau n+\bar\nu l,\cr
&\delta l=
(\beta+\bar\alpha)l-\bar\rho m-\sigma\bar m\hbox{,~~~}
\delta m=
(\beta-\bar\alpha)m-\sigma n+\bar\lambda l,\cr
&\bar\delta l=
(\alpha+\bar\beta)l-\bar\sigma m-\rho\bar m\hbox{,~~~}
\bar\delta m=
(\alpha-\bar\beta)m-\rho n+\bar\mu l.
}
\endeq

For the Kerr--de Sitter metric some of the connection components
vanish,
\begin
\kappa=\sigma=\lambda=\nu=0.
\endeq
The vanishing of these connection components is a feature common to all
black hole solutions. The remaining connection components are
\begin
\eqalign{
&\rho=\mu=-{1\over\sqrt{2}}{c\over\rho^3}\sqrt{\Delta_r}
\hbox{,~~~~}
\epsilon=\gamma=-{1\over2\sqrt{2}}{c\over\rho^3}\sqrt{\Delta_r}
+{1\over2\sqrt{2}}{\sqrt{\Delta_r}'\over\rho},\cr
&\tau=-\pi={i\over\sqrt{2}}{c\over\rho^3}\sqrt{\Delta_\mu}
\hbox{,~~~~}
\beta=-\alpha={i\over2\sqrt{2}}{c\over\rho^3}\sqrt{\Delta_\mu}
+{1\over2\sqrt{2}}{\sqrt{\Delta_\mu}'\over\rho},
}\endeq
where
\begin
c=r+i\mu.
\endeq

The electromagnetic field is described by three complex scalars
$\phi_0$, $\phi_1$ and $\phi_2$, with the radial fields in $\phi_1$ and
the axial fields in $\phi_0+\phi_2$. They satisfy Maxwell's equations,
\begin
\eqalign{
&D\phi_1-\bar\delta\phi_0=
(\pi-2\alpha)\phi_0+2\rho\phi_1-\kappa\phi_2,\cr
&\delta\phi_1-\Delta\phi_0=
(\mu-2\gamma)\phi_0+2\tau\phi_1-\sigma\phi_2,\cr
&D\phi_2-\bar\delta\phi_1=
(\rho-2\epsilon)\phi_2+2\pi\phi_1-\lambda\phi_0,\cr
&\delta\phi_2-\Delta\phi_1=
(\tau-2\beta)\phi_2+2\mu\phi_1-\nu\phi_0.
}\endeq

We follow reference \cite{1} and introduce new radial and angular
derivatives defined by
\begin
{\cal D}_n=\partial_r+{iK_r\over \Delta_r}+n{\Delta_r'\over\Delta_r}
\hbox{,~~~~~}
{\cal D}^\dagger_n=\partial_r-{iK_r\over \Delta_r},
+n{\Delta_r'\over\Delta_r},
\endeq
and
\begin
{\cal L}_n=\partial_\mu+{K_\mu\over
\Delta_\mu}+n{\Delta_\mu'\over\Delta_\mu}
\hbox{,~~~~~}
{\cal L}^\dagger_n=\partial_\mu-{K_\mu\over \Delta_\mu}
+n{\Delta_\mu'\over\Delta_\mu},
\endeq
with $K_r$ and $K_\mu$ defined in equations \ref{3.14} and \ref{3.6}.
These are respectively purely radial or angular in character. Useful
properties of these derivatives are that
\begin
{\cal D}_n+{n\over\bar c}=
(\bar c)^{-n}{\cal D}_n(\bar c)^n\hbox{,~~~~}
{\cal D}_n=\Delta_r^{-n}{\cal D}_0\Delta_r^n
\endeq
and similarly for ${\cal L}$. For scalar functions that have the
dependence $\exp i(m\phi-\omega t)$ on $\phi$ and $t$ it is possible to
write the directional derivatives along the tetrad vectors as
\begin
D=\left({\Delta_r\over2\rho^2}\right)^{1/2}{\cal D}_0\hbox{,~~~~}
\Delta=-\left({\Delta_r\over2\rho^2}\right)^{1/2}{\cal D}^\dagger_0
\hbox{,~~~~}
\delta=\left({\Delta_\mu\over2\rho^2}\right)^{1/2}{\cal L}_0.
\endeq

We can now write Maxwell's equations in terms of ${\cal D}$ and ${\cal
L}$, using the connection components from equation \ref{4.6}. We will
rescale the Maxwell fields first, to $\varphi_i=\bar c\phi_i$, and then
the equations become
\begin
\eqalign{
\sqrt{\Delta_r}\left({\cal D}_0+{1\over\bar c}\right)\varphi_1=
&\sqrt{\Delta_\mu}
\left({\cal L}^\dagger_{1/2}+{i\over\bar c}\right)\varphi_0\cr
-\sqrt{\Delta_r}\left({\cal D}^\dagger_{1/2}-{1\over\bar
c}\right)\varphi_0=
&\sqrt{\Delta_\mu}
\left({\cal L}_0-{i\over\bar c}\right)\varphi_1\cr
\sqrt{\Delta_r}\left({\cal D}_{1/2}-{1\over\bar c}\right)\varphi_2=
&\sqrt{\Delta_\mu}
\left({\cal L}^\dagger_0-{i\over\bar c}\right)\varphi_1\cr
-\sqrt{\Delta_r}\left({\cal D}^\dagger_0+{1\over\bar
c}\right)\varphi_1=
&\sqrt{\Delta_\mu}
\left({\cal L}_{1/2}+{i\over\bar c}\right)\varphi_2.
}\endeq

The field $\varphi_1$ may be eliminated from \ref{4.13},
\begin
\sqrt{\Delta_\mu}
\left({\cal L}_0-{i\over\bar c}\right)
\sqrt{\Delta_\mu}
\left({\cal L}^\dagger_{1/2}+{i\over\bar c}\right)\varphi_0=
-\sqrt{\Delta_r}\left({\cal D}_0+{1\over\bar c}\right)
\sqrt{\Delta_r}\left({\cal D}^\dagger_{1/2}-{1\over\bar
c}\right)\varphi_0
\endeq
{}From the forms of ${\cal D}$ and ${\cal L}$, and of $K_r$ and $K_\mu$,
this can be simplified to
\begin
\left({\cal L}_{-1/2}\Delta_\mu{\cal L}^\dagger_{1/2}
+{\cal D}_{-1/2}\Delta_r{\cal D}^\dagger_{1/2}
+2i\omega\chi^2c\right)\varphi_0=0.
\endeq
Similarly,
\begin
\left({\cal L}^\dagger_{-1/2}\Delta_\mu{\cal L}_{1/2}
+{\cal D}^\dagger_{-1/2}\Delta_r{\cal D}_{1/2}
-2i\omega\chi^2c\right)\varphi_2=0.
\endeq
This is the complex conjugate of the equation for $\varphi_0$. The
result for $\varphi_1$ can be obtained by subtracting the middle two
equations in \ref{4.13},
\begin
K_\mu\phi_1=-\case1/2\sqrt{\Delta_\mu\Delta_r}
\left({\cal D}^\dagger_{1/2}\phi_0+{\cal D}_{1/2}\phi_2\right).
\endeq

The equations separate in the same way as the scalar field equation,
that is we set
\begin
\varphi_0={\cal R}_0(r){\cal S}_0(\mu)e^{im\phi}e^{-i\omega t}.
\endeq
Then,
\begin
\eqalign{
\left({\cal L}_{-1/2}\Delta_\mu{\cal L}^\dagger_{1/2}
-2\omega\chi^2\mu\right){\cal S}_0&=-\lambda{\cal S}_0\cr
\left({\cal D}_{-1/2}\Delta_r{\cal D}^\dagger_{1/2}
+2i\omega\chi^2r\right){\cal R}_0&=\lambda{\cal R}_0.
}\endeq
The second of these equations is the analogue for Kerr--de Sitter of
Teukolsky's equation for Kerr\cite{13}. For spin $s$ we would have,
\begin
\left({\cal D}_{-s/2}\Delta_r{\cal D}^\dagger_{s/2}
+2(2s-1)i\omega\chi^2r\right){\cal R}=\lambda{\cal R}.
\endeq

We would like to express the radial equation in terms of the $r^*$
coordinate as we did for the scalar field. Before doing this it
is advantageous to introduce electromagnetic potentials rather
than the fields themselves because these are directly comparable to
the scalar field, particularly when we come to evaluate energy fluxes.
In order to proceed we need the Teukolsky and Starobinsky identities,
\begin
{\cal D}_{-1/2}\Delta_r {\cal D}_{1/2}{\cal R}_2
={\cal C}{\cal R}_0\hbox{,~~~}
{\cal D}^\dagger_{-1/2}\Delta_\mu {\cal D}^\dagger_{1/2}{\cal R}_0
={\cal C}{\cal R}_2,
\endeq
where
\begin
{\cal C}^2=\lambda^2+4\omega(am-a^2\omega)\chi^4.
\endeq
There are similar identities for ${\cal L}$. The derivation of these
identities follows the corresponding results in reference \cite{1}.

Define
\begin
R_0={\cal D}\Delta_r^{1/2}{\cal R}_2\hbox{,~~~~}
R_2={\cal D}^\dagger\Delta_r^{1/2}{\cal R}_0.
\endeq
It follows from the identities that
\begin
{\cal R}_0={\cal C}^{-1}\Delta_r^{1/2}{\cal D}R_0\hbox{,~~~~}
{\cal R}_2={\cal C}^{-1}\Delta_r^{1/2}{\cal D}^\dagger R_2.
\endeq

Combining the two sets of radial equations gives
\begin
{\cal D}^\dagger\Delta_r{\cal D}R_0={\cal C}R_2\hbox{,~~~~}
{\cal D}\Delta_r{\cal D}^\dagger R_2={\cal C}R_0.
\endeq
These expand into a form,
\begin
\eqalign{
\left(\partial_r\Delta_r\partial_r+K_r^2/\Delta_r
-2i\omega\chi^2r\right)R_0&={\cal C} R_2\cr
\left(\partial_r\Delta_r\partial_r+K_r^2/\Delta_r
+2i\omega\chi^2r\right)R_2&={\cal C} R_0
}
\endeq
This coupled form of the equations seems a step backwards, but it
will still be acceptable for our analysis.
The equations can be written as a scattering problem,
\begin
\left({d^2\over d r^{*2}}+V_s(r)\right)\sigma_r F=0
\endeq
where
\begin
F=\pmatrix{R_0+R_2\cr  R_0- R_2}
\hbox{,~~~~} V=\pmatrix{V_+&U\cr-U&V_-},
\endeq
and
\begin
V_\pm=(\omega-m\Omega)^2
\mp{{\cal C}\Delta_r\over\chi^4\sigma_r^4}
-{\Delta_r\over\chi^4\sigma_r^3}\left({r\Delta_r\over\sigma_r^3}\right)'
\hbox{,~~~~~}
U=-{2i\omega r\Delta_r\over\chi^2\sigma_r^4}.
\endeq
It would be possible to separate these equation using the methods
described in \cite{1}, but this will not be necessary.

The new radial functions can be used to define new fields,
\begin
\Phi_0=R_0(r){\cal S}_0(\mu)e^{im\phi}e^{-i\omega t}.
\endeq
These are related to the electromagnetic potentials. The result
\ref{4.24} shows that the Maxwell fields are given by
\begin
\phi_0={\sqrt{\Delta_r}\over {\cal C}\,\bar c}{\cal
D}\,\Phi_0\hbox{,~~~~}
\phi_2={\sqrt{\Delta_r}\over {\cal C}\,\bar c}{\cal D}^\dagger\Phi_2,
\endeq
and from \ref{4.17},
\begin
\phi_1=
{\sqrt{\Delta_r\Delta_\mu}\over 2K_\mu}(\phi_0+\phi_1)
-{\sqrt{\Delta_\mu}\over 2K_\mu}(\Phi_0+\Phi_2).
\endeq
We have not fully explored the numerous relationships that exist
between the mode functions because they are not needed for our
stability analysis, but there are many equivalent ways to express the
above results.

\section{The gravitational wave equations}

In the Newman--Penrose formalism, the Weyl tensor is described by five
complex scalars $\Psi_0\dots\Psi_4$. Of these, only $\Psi_2$ is
non--vanishing in black hole solutions. The remaining scalars can be
used to describe gravitational wave perturbations of the metric. In
this section we will separate the equations for these perturbations on
the Kerr--de Sitter background. The perturbed metric has other degrees
of freedom, but we leave those equations for another occasion.

The Bianchi identities for the Weyl tensor components read
\begin
\eqalign{
(\bar\delta-4\alpha+\pi)\Psi_0-(D-2\epsilon-4\rho)\Psi_1
&=3\kappa\Psi_2\cr
(\Delta-4\gamma+\mu)\Psi_0-(\delta-4\tau-2\beta)\Psi_1
&=3\sigma\Psi_2\cr
(\delta+4\beta-\tau)\Psi_4-(\Delta+2\gamma+4\mu)\Psi_3
&=-3\nu\Psi_2\cr
(D+4\epsilon-\rho)\Psi_4-(\bar\delta+4\pi+2\alpha)\Psi_3
&=-3\lambda\Psi_2.
}\endeq
The connection components $\kappa$, $\sigma$, $\lambda$ and $\nu$ all
vanish for the unperturbed metric and the values appearing here are
therefore the perturbed values. We need two Ricci identities to solve
for these,
\begin
\eqalign{
(D-\rho-\bar\rho-3\epsilon+\bar\epsilon)\sigma
-(\delta-\tau+\bar\pi-\bar\alpha-3\beta)\kappa=\Psi_0\cr
(\Delta-\mu-\bar\mu+3\gamma-\bar\gamma)\lambda
-(\bar\delta+3\alpha+\bar\beta+\pi-\bar\tau)\nu
=-\Psi_4.
}\endeq

The angular and radial operators ${\cal L}$ and ${\cal D}$ that where
used for the Maxwell fields can be used here also. We rescale the Weyl
scalars first, $\psi_i=c^2\Psi_i$ and the connection components
$\hat\kappa=c^{1/2}(\bar c)^{-1/2}\kappa$. Then,
\begin
\eqalign{
\sqrt{\Delta_\mu}
\left({\cal L}_1^\dagger+{3i\over\bar c}\right)\psi_0
-\sqrt{\Delta_r}
\left({\cal D}_{-1/2}+{3\over\bar c}\right)\psi_1
=3\sqrt{2}(\bar c\psi_2)\hat\kappa,\cr
\sqrt{\Delta_\mu}
\left({\cal L}_{-1/2}-{3i\over\bar c}\right)\psi_1
+\sqrt{\Delta_r}
\left({\cal D}^\dagger_1-{3\over\bar c}\right)\psi_0
=-3\sqrt{2}(\bar c\psi_2)\hat\sigma,\cr
\sqrt{\Delta_\mu}
\left({\cal L}_1+{3i\over\bar c}\right)\psi_4
+\sqrt{\Delta_r}
\left({\cal D}^\dagger_{-1/2}+{3\over\bar c}\right)\psi_3
=-3\sqrt{2}(\bar c\psi_2)\hat\nu,\cr
\sqrt{\Delta_\mu}
\left({\cal L}^\dagger_{-1/2}-{3i\over\bar c}\right)\psi_3
-\sqrt{\Delta_r}
\left({\cal D}_1-{3\over\bar c}\right)\psi_4
=3\sqrt{2}(\bar c\psi_2)\hat\lambda.
}\endeq
and
\begin
\eqalign{
\sqrt{\Delta_\mu}
\left({\cal L}_{-1/2}-{3i\over\bar c}\right)\hat\kappa
-\sqrt{\Delta_r}
\left({\cal D}_{-1/2}+{3\over\bar c}\right)\hat\sigma
=-\sqrt{2}c(\bar c)^{-2}\psi_0,\cr
\sqrt{\Delta_\mu}
\left({\cal L}^\dagger_{-1/2}-{3i\over\bar c}\right)\hat\nu
+\sqrt{\Delta_r}
\left({\cal D}^\dagger_{-1/2}+{3\over\bar c}\right)\hat\lambda
=\sqrt{2}c(\bar c)^{-2}\psi_4.
}\endeq

The Weyl scalar
\begin
\Psi_2=-{M\over(\bar c)^3}-{Q^2c\over(\bar c)^3}.
\endeq
If the charge $Q=0$, we can eliminate $\psi_1$ from the first two
equations by using equation \ref{5.4}. We will restrict our attention
to this case alone, then
\begin
\eqalign{
\sqrt{\Delta_\mu}
\left({\cal L}_{-1/2}-{3i\over\bar c}\right)
\sqrt{\Delta_\mu}
\left({\cal L}_1^\dagger+{3i\over\bar c}\right)\psi_0
+\sqrt{\Delta_r}
\left({\cal D}_{-1/2}+{3\over\bar c}\right)
\sqrt{\Delta_r}
&\left({\cal D}^\dagger_1-{3\over\bar c}\right)\psi_0\cr
&=6{c M\over (\bar c)^2}\psi_0.}
\endeq
After simplification, this gives
\begin
\left({\cal L}_{-1}\Delta_\mu{\cal L}_1^\dagger
+{\cal D}_{-1}\Delta_r{\cal D}_1^\dagger+6i\omega\chi^2c
-2\Lambda\rho^2\right)\psi_0=0.
\endeq
Similarly, the Weyl scalar $\psi_4$ satisfies the complex conjugate
equation. The remaining field $\psi_3$ is gauge dependent and can be
made to satisfy similar equations or made to vanish, depending on the
choice of tetrad in the perturbed metric.

We separate the equation as before with a radial function ${\cal
R}_0(r)$ and an angular function ${\cal S}_0(\mu)$. These functions
satisfy
\begin
\eqalign{
\left({\cal L}_{-1}\Delta_\mu{\cal L}^\dagger_{1}
-6\omega\chi^2\mu-2\Lambda\mu^2\right){\cal S}_0
&=-\lambda {\cal S}_0\cr
\left({\cal D}_{-1}\Delta_r{\cal D}^\dagger_{1}
+6i\omega\chi^2r-2\Lambda r^2\right){\cal R}_0
&=\lambda {\cal R}_0.
}\endeq
These equations are consistent with the new version of Teukolsky's
equation which was proposed in the previous section, but with the
addition of a cosmological constant term. This term plays a similar
role to a mass in the case of a massive scalar.

As before, we will introduce potentials to facilitate the stability
analysis. This makes use of the identities,
\begin
{\cal D}_{-1}\Delta{\cal D}{\cal D}\Delta{\cal D}_1{\cal R}_4
={\cal C}{\cal R}_0\hbox{,~~~~}
{\cal D}^\dagger_{-1}\Delta{\cal D}^\dagger{\cal D}^\dagger
\Delta{\cal D}^\dagger_1{\cal R}_0
=\bar{\cal C}{\cal R}_4.
\endeq
The modulus of the constant ${\cal C}$ is given by
\begin
\eqalign{
|\,{\cal C}|^2=&\lambda^2\lambda_1^2-20K_0K_2\lambda\lambda_1
+24K_2^2a^2\lambda+K_2(K_0^2+M^2)\cr
&+8\Lambda(a^2\lambda\lambda_1-4K_0^2\lambda+6a^2K_0K_2)+16\Lambda^2a^4
}\endeq
where
\begin
K_0=(am-a^2\omega)\chi^2\hbox{,~~~~}K_2=-2\omega\chi^2
\hbox{,~~~~}\lambda_1=\lambda+2-\case2/3\Lambda a^2.
\endeq

Define
\begin
R_0={\cal D}{\cal D}\Delta{\cal D}_1{\cal R}_4\hbox{,~~~~}
R_4={\cal D}^\dagger{\cal D}^\dagger
\Delta{\cal D}^\dagger_1{\cal R}_0,
\endeq
and also
\begin
\Phi_i=R_i(r)S_i(\theta)e^{im\phi}e^{-i\omega t}.
\endeq
The equations for these radial functions form a coupled set of second
order equations. (This can be shown by repeated use of the original
equation to reduce the order, but the resulting equations are quite
complicated). The Weyl curvature scalars are recoverable from,
\begin
\Psi_4={\Delta\over \bar{\cal C}\,c^2}{\cal
D}^\dagger\Phi_4\hbox{,~~~~}
\Psi_0={\Delta\over {\cal C}\,c^2}{\cal D}\,\Phi_0.
\endeq
The remaining scalars $\Psi_1$ and $\Psi_3$ can be chosen to be zero by
adopting a suitable gauge.
\vfill\eject

\section{Regularity at the Cauchy horizon}

We shall now examine the behaviour of an incoming flux of energy as it
reaches the Cauchy horizon. The energy flux will be set up initially in
the form of waves that enter the black hole from the outside. These
waves are affected by  the geometry of spacetime in two ways, their
wavelengths are stretched or compressed and they get scattered. It is
the compression of the waves that leads to instability, but it is
necessary to consider the scattering problem in some detail. We begin
therefore with some general remarks about scattering problems. The
following treatment is based very closely on Chandrasekhar and Hartle
\cite{16}.

Consider the following scattering problem
\begin
{d^2f\over d x^2}+V(x)f=0
\endeq
where the potential has an exponential expansion as $x\to\pm\infty$,
\begin
V(x)=\left\{\matrix{\sum_{n=0}^\infty c_n^+
e^{-2n\kappa_+x}&x\to\infty\cr
\sum_{n=0}^\infty c_n^- e^{2n\kappa_-x}&x\to-\infty\cr}\right..
\endeq
Suppose also that $c_0^\pm$ is parameterised by a frequency $\omega$,
\begin
c_0^\pm=(\omega-m\Omega_\pm)^2.
\endeq
A potential of this form is called Yukawian \cite{17} and leads to
special analyticity properties of the scattering coefficients which are
derived in the appendix.

In the case of waves entering the black hole there are two scattering
regions, the first outside the hole in region I and the second inside
region II. The radial modes will be separated into $\leftarrow$ modes
which enter from the cosmological horizon and $\rightarrow$ modes that
are scattered from the past event horizon. These are shown in figure 5.
Near to the Cauchy horizon, the asymptotic limits of the modes are
\begin
\sigma_rR\sim Ae^{ik_3r^*}+Be^{-ik_3r^*}
\endeq
in both cases.

The scattering amplitudes can be found by combining two scattering
problems with coordinates $x=r^*$ in both regions. This gives a
potential of the required form. The scattering problems are related
to those in the appendix by the substitutions
\begin
\eqalign{
{\rm I}:\ &r_-=r_2,\ r_+=r_1,
\ f_2=\sigma_r\overleftarrow R,
\ f_1=\sigma_r\overrightarrow R;\cr
{\rm II}:\ &r_-=r_2,\ r_+=r_3,
\ f_2=(T_{II}/T_I)\sigma_r\overleftarrow R,
\ f_1=(T_{II}/R_I)\sigma_r\overrightarrow R.\cr
}
\endeq

Consequently,
\begin
\overleftarrow A={T_I\over T_{II}}
\hbox{,~~~~~}
\overrightarrow A={R_I\over T_{II}}.
\endeq

The full expansion of the field $\Phi$ will be of the form
$\Phi=\overleftarrow\Phi+\overrightarrow\Phi$, with
\begin
\overleftarrow\Phi=
\sum_{l,m}\int{d\omega\over2\pi}
\overleftarrow W(lm\omega)
\overleftarrow R(\omega;r)
S(lm\omega;\theta)
e^{im\phi-i\omega t}.
\endeq
The functions $W(lm\omega)$ determine the initial conditions on an
initial surface in region I. We will take $W$ to be a regular function
of $\omega$ with isolated poles.

Close to the Cauchy horizon, the asymptotic form of the mode functions
can be used, and then
\begin
\overleftarrow\Phi=
\sum_{l,m}\int{d\omega\over2\pi\sigma_r}
\overleftarrow W(lm\omega)
S(lm\omega;\theta)
(\overleftarrow Ae^{-ik_3 v}
+\overleftarrow Be^{ik_3 u})\,e^{im(\phi-\Omega t)}.
\endeq
Recall that $\phi-\Omega(r_3)\,t$ is the regular coordinate $\phi_3$.
Futhermore,  $k_3$ is taken to be real or in the lower half plane and
therefore the exponential terms converge at the Cauchy horizon, where
$v\to\infty$. In fact, the only possible divergences in the field
$\Phi$ at the horizon would arise from poles in $A(\omega)$, and these
can be removed by taking the principal value of the integral.

The energy flux in the fields would be measured by the components of
the stress--energy tensor. A coordinate frame that is regular on the
Cauchy horizon, such as the Kruskal coordinate frame, should be used. A
typical component of the stress--energy for a scalar field would be
\begin
T_{VV}=-\partial_V\Phi\,\partial_V\Phi.
\endeq
The part of the stress--energy tensor that is able to diverge at the
Cauchy horizon is the derivative with respect to $V$. This is because
\begin
\partial_V\Phi=e^{\kappa_3v}\partial_v\Phi.
\endeq
At the Cauchy horizon, $v\to\infty$, and the exponential diverges.

For the electromagnetic field, the stress tensor components in the null
tetrad frame are functions of the Maxwell fields $\phi_0$, $\phi_1$ and
$\phi_2$, for example
\begin
T_{ab}l^al^b=\phi_0\bar\phi_0\hbox{,~~~}
T_{ab}n^an^b=\phi_2\bar\phi_2\hbox{,~~~}
T_{ab}l^an^b=\phi_1\bar\phi_1.
\endeq
This frame has to be rescaled to obtain the Kruskal frame,
\begin
\partial_V= e^{\kappa_3(u+v)/2}\,l\hbox{,~~~}
\partial_U= e^{-\kappa_3(u+v)/2}\,n.
\endeq
The potentially diverging stress--energy tensor component is then
\begin
T_{VV}={2\rho^2\over{\cal C}^2{\bar c}^2}
(\partial_V\Phi_0)^*(\partial_V\Phi_0),
\endeq
after using expresions for the Maxwell fields given in equations
\ref{4.31}.

In the case of the gravitational waves we are interested in whether
components of the curvature tensor remain finite at the Cauchy horizon.
We have
\begin
C_{abcd}l^am^bl^cm^d=-\Psi_0\hbox{,~~~~~}
C_{abcd}n^a\bar m^bn^c\bar m^d=-\Psi_4.
\endeq
Again, the important terms are the ones depending upon the $V$
derivatives of $\Phi$. (The basis vectors $m$ and $\bar m$ have to be
Lorentz boosted to a regular frame.)

Using the mode decomposition above and the asymptotic forms of the mode
functions we can express the potentially divergent term as
\begin
\partial_V\overleftarrow\Phi=
ie^{\kappa_3v}
\sum_{l,m}\int{d\omega\over2\pi\sigma_r}
\overleftarrow W(lm\omega)
k_3\overleftarrow A(\omega)
S(lm\omega;\theta)
e^{im\phi_3}e^{-ik_3 v}.
\endeq
and a similar term with the $\rightarrow$ amplitude. The integrand is
regular with isolated poles which allows us to move the contour of
integration away from the real axis. The behaviour of the whole term as
$v\to\infty$ depends upon the poles in the integrand in the lower half
of the complex $\omega$ plane.

The first such pole in $k_3A$ can be found from equation \ref{8.11} and
figure 6. The crucial feature is that the pole at
$\omega=m\Omega_2-i\kappa_2$ in $1/T_{II}$ is matched by an identical pole
in $1/T_I$ or $R_I/T_I$. The ratio \ref{6.6} therefore has no pole there.
The pole at $\omega=m\Omega_3$ also cancels with $k_3$,
leaving the first pole at $\omega=m\Omega_3-i\kappa_3$.

In order to discuss the behavior of $W$ we can consider in a similar
way the stress--energy tensor component $T_{VV}$ at the cosmological
horizon. Here $\kappa_3$ is replaced by $\kappa_1$, and we see that the
value of $T_{VV}$ on the horizon is determined by the residue of $W$ at
$\omega=m\Omega_1-i\kappa_1$. Therefore, if the stress energy tensor is
non--zero at the cosmological horizon, there should be a pole in $W$
with imaginary part $-i\kappa_1$.

We have stated already our assumption that the angular functions are
analytic in a strip bordering the real axis. We conclude that it is
possible to distort the contour of integration to pass parallel to the
real axis through $-i\kappa_1$ but no further, and that $T_{VV}$ at the
Cauchy horizon diverges as $\exp(\kappa_3-\kappa_1)v$. In particular,
the stress--energy tensor of the waves is regular at the Cauchy horizon
if the surface gravity there is less than the surface gravity at the
cosmological horizon.

\section{Conclusion}

We have seen that there are situations in which the energy fluxes of
scalar, electromagnetic and gravity waves remain finite at the Cauchy
horizon and allow an observer to pass safely through into the interior
of the black hole. The angular velocity of the hole has to be finely
tuned to allow this to happen, and of course we assumed the existence
of a cosmological constant.

The separation of the field equations that enabled us to make progress
was closely associated with the algebraic character of the curvature
tensor of the black hole and with the symmetrical appearance of the
metric with respect to the angular and radial coordinates. This also
explains why the angular and radial equations have such similar
properties \cite{1}. We have been able to predict the form of the
radial field equations with a cosmological constant for arbitrary spin,
thus generalising Teukolsky's results \cite{13}. We have also found it
necessary to assume some properties of the angular mode functions that
merit further study.

The gravitational wave equations form a major part of the full stability
analysis for this spacetime. We have left the complete separation of the
gravitational perturbations in the metric for another time. The remaining
metric components are mostly gauge parameters of little physical interest
in the absence of other fields. The more interesting case would include
electromagnetic fields as well, but separation is then problematic,
even in the standard analysis, particularly when the hole carries an
electric charge.

Even in the case that the cosmological constant is very small, there is
still a tiny range of values for the rotation rate of the hole leading
to a stable Cauchy horizon. This range of values is very close to the
extreme limit in which the surface gravity of the event horizon vanishes,
a situation that we would expect only to be attainable asymptotically. It
therefore seems possible that attempts to spin up a black hole could be
used in principle to stabilise the Cauchy horizon.

The stable situation raises the question of how to describe a naked
singularity. The classical theory needs to be modified to do this, but
we might make some progress by treating the interior geometry as a
central force problem with an undetermined set of phase shifts at the
singularity. The next question would be what are the consequences of
closed timelike curves in the interior of the hole. There is a
possibility that linear wave propagation and therefore the stability
analysis could still be consistently defined, as they are in
asymtotically flat spacetimes \cite{18}. Finally, we would be left with
the continuation of the observer's journey through the hole and into
another universe.

\section{Appendix}

In this appendix we describe the procedure by which the poles of the
scattering amplitudes for equation \ref{6.1} can be found, following
ref \cite{16} but allowing the potential to approach different constant
values at the horizons. Define solutions of the scattering problem
$f_1$ and $f_2$ with the asymptotic behaviour
\begin
\eqalign{
f_1(x,\omega)&\to e^{+ik_+x}\hbox{~~~~}
\bar f_1(x,\omega)\to e^{-ik_+x}\hbox{~~~~~}x\to\infty\cr
f_2(x,\omega)&\to e^{-ik_-x}\hbox{~~~~}
\bar f_2(x,\omega)\to e^{+ik_-x}\hbox{~~~~~}x\to-\infty
}
\endeq
where
\begin
k_\pm=\omega-m\Omega_\pm.
\endeq
(Note that $\pm$ refer to $\pm\infty$, the reverse of ref. \cite{16}.)

The solutions $f_1$ and $\bar f_1$ form a complete set of solutions to
the scattering problem and it is therefore possible to express $f_2$ as
a linear combination,
\begin
f_2(x,\omega)={R(\omega)\over T(\omega)}f_1(x,\omega)
+{1\over T(\omega)}\bar f_1(x,\omega).
\endeq
This defines the reflection and transmission coefficients $R(\omega)$
and $T(\omega)$. They can be extracted from $f_1$ and $f_2$ for any
value of $x$ by using the Wronskians,
\begin
\eqalign{
{1\over T(\omega)}&=-{1\over2ik_+}[f_1(x,\omega),f_2(x,\omega)]\cr
{R(\omega)\over T(\omega)}&=+{1\over2ik_+}[\bar
f_1(x,\omega),f_2(x,\omega)]
}.
\endeq
where $[f,g]=fg'-gf'$.

The scattering problem can be reduced to an integral equation by using
the Green function. We first write the differential equation as
\begin
-{d^2f_2\over d x^2}+ k_-^2f_2=\left(V(x)-c_0^-\right)f_2.
\endeq
The solution obtained by regarding the right hand side as a source term
is
\begin
f_2(x,\omega)=e^{-ik_-x}+\int_{-\infty}^x {1\over k_-}
\sin k_-(x-x')\left(V(x')-c_0^-\right)f_2(x',\omega)\,dx'.
\endeq

The integral equation can be iterated to obtain a series for $f_2$. set
\begin
f_2(x,\omega)=e^{-ik_-x}+\sum_{s=1}^\infty f^{(s)}_2(x,\omega)
\endeq
then,
\begin
f_2^{(s)}(x,\omega)=\int_{-\infty}^x {1\over k_-}
\sin k_-(x-x_s)\left(V(x_s)-c_0^-\right)f_2^{(s-1)}(x_s,\omega)\,dx_s.
\endeq
Repeatedly iterating this equation gives
\begin
f_2^{(s)}(x,\omega)={e^{-ik_- x}\over(2ik_-)^s}
\int_{-\infty}^{x_0} dx_0\dots\int_{-\infty}^{x_s} dx_s
\prod_{i=1}^s \left\{\left(e^{2ik_- (x_{i-1}-x_i)}-1\right)
\sum_{n=1}^\infty c_n^-e^{2n\kappa_-x_i}\right\}
\endeq
where the exponential expansion of $V$ has been inserted.
If we define $y_i=x_{i-1}-x_i$, then following \cite{16}, we can write
\begin
f_2^{(s)}(x,\omega)={e^{-ik_- x}\over(2ik_-)^s}
\int_0^\infty dy_1\dots\int_0^\infty dy_s
\prod_{i=1}^s \left\{(1-e^{2ik_- y_i})
\sum_{n=1}^\infty c_n^-e^{2n\kappa_-x_i}\right\}.
\endeq

The integrand is a sum of exponentials and the result of integrating
over the $y_i$ introduces poles at $k_-=-in\kappa_-$, where
$n=1,2,\dots$. Therefore, $f_2(x,\omega)$ has poles at
\begin
\omega=m\Omega_--in\kappa_-.
\endeq
A similar treatment of $f_1(x,\omega)$ shows that this function has its
poles at \begin
\omega=m\Omega_++in\kappa_+.
\endeq
{}From equation \ref{8.4}, these determine where the poles of
$R(\omega)/T(\omega)$ and $1/T(\omega)$ lie, as shown in figure 6.

\vfill\eject
\noindent References
\medskip
\item{1.}S. Chandrasekhar, The Mathematical Theory of Black Holes, (New
York, Oxford University Press, 1983).
\item{2.}W. Israel, in Black Hole Physics, ed V. De Sabbata and Z.
Zhang (Amsterdam: Kluwer Academic Publishers 1992).
\item{3.}F. Mellor and I. G. Moss, Phys. Rev. D41, 403 (1990).
\item{4.}F. Mellor and I. G. Moss, Class. Quantum Grav. 9, L43 (1992).
\item{5.}P. R. Brady and E. Poisson, Class. Quantum Grav. 9, 121
(1992).
\item{6.}P. R. Brady, D. Nunez and S. Sinha, Phys. Rev. D47, 4239
(1993).
\item{7.}B. Carter, in Black Holes ed C. DeWitt and B. S. DeWitt (New
York, Gordon and Breach 1973).
\item{8.}F. Mellor and I. G. Moss, Class. Quantum Grav. 6, 1379
(1989).
\item{9.}G. W. Gibbons and S. W. Hawking, Phys. Rev. D15, 2738 (1977).
\item{10.}B. Carter, Comm. Math. Phys. 10, 280 (1968).
\item{11.}J. Meixner, F. W. Sch\"afke and G. Wolf, Mathieu Functions,
Spheroidal Functions and their Mathematical Foundations (Heidelberg,
Springer--Verlag 1980).
\item{12.}E. T. Newman and R. Penrose, J. Math. Phys. 3, 566 (1962).
\item{13.}S. A. Teukolsky, Phys. Rev. Lett 29, 1114 (1972).
\item{14.}A. A. Starobinsky and S. M. Churilov, Soviet Phys. JETP 38, 1
(1973).
\item{15.}R. Geroch, A. Held and R. Penrose, J. Math. Phys. 14, 874
(1973).
\item{16.}S. Chandrasekhar and J. B. Hartle, Proc. R. Soc. A384, 301
(1982).
\item{17.}V. de Alfaro and T. Regge, Potential Scattering (Amsterdam,
North Holland 1965).
\item{18.}J. Friedman, M. S. Morris, I. D. Novikov, F. Echeverria, G.
Klinkhammer, K. S. Thorne and U. Yurtsever, Phys. Rev. D42 1915
(1990).
\vfill
\eject
\noindent Figures
\medskip
\item{1.}The parameter space of the black hole for (a) $Q=0$ and
(b) $Q=0.2$.
The mass $M$ and rotation rate $a$ have both been scaled by $\alpha$,
where $\alpha^2=3/\Lambda$.
Coincident horizons occur along the outer borders of the triangular
region, $r_2=r_3$ on OA and $r_1=r_2$ on AC. Corresponding surface
gravities also vanish there.
Interior lines denote the cases where $\kappa_1=\kappa_3$ (the lower
line OA) and $\kappa_1=\kappa_2$ (along OB). The narrow region extending
from O to A is the stability region.
\item{2.}A Penrose diagram showing the event horizon $r_2$
between regions I (outside the hole) and II (inside the hole).
\item{3.}Part of the Penrose diagram of the fully extended spacetime.
Horizon $1$ is a cosmological de Sitter horizon, horizon $2$ a black
hole event horizon and horizon $3$ an inner Cauchy horizon. Asymptotic
regions exist and are labelled as ${\cal J}$. Singularities at $r=0$ are
represented by zigzag lines. The fully extended diagram covers the
whole plane, including the (blank) regions that can only be reached
by passing through $r=0$.
\item{4.}The radial potential (reversed sign) in region II for various
values of $l$. The upper three curves have $m=1$ and the lower three
$m=0$.
\item{5.}The inward and outward radial mode functions $\overleftarrow
R$ and $\overrightarrow R$ showing how their asymptotic amplitudes
$A$, $B$ and $C$ are associated with the respective horizons.
\item{6.}The location of the poles in $1/T$ and $R/T$ for $s=0$ are
shown in the complex $\omega$--plane.
\vfill
\eject
\noindent Tables
\medskip
\bigskip
\hfil\vbox{\offinterlineskip
\hrule
\halign{&\vrule#&\strut\quad\hfil$#$\quad\cr
height2pt&\omit&&\omit&&\omit&&\omit&&\omit&&\omit&\cr
&\omega\,a&&(l,m)=(0,0)&&(1,0)&&(1,1)&&(1,-1)&&(2,0)&\cr
height2pt&\omit&&\omit&&\omit&&\omit&&\omit&&\omit&\cr
\noalign{\hrule}
height2pt&\omit&&\omit&&\omit&&\omit&&\omit&&\omit&\cr
&0.0&&0.000&&2.000&&2.000&&2.000&&6.000&\cr
&0.1&&0.007&&2.004&&1.809&&2.208&&6.005&\cr
&0.2&&0.027&&2.016&&1.631&&2.432&&6.019&\cr
&0.3&&0.060&&2.036&&1.471&&2.672&&6.043&\cr
&0.4&&0.107&&2.064&&1.327&&2.982&&6.076&\cr
&1.0&&0.667&&2.400&&0.800&&4.800&&6.476&\cr
height2pt&\omit&&\omit&&\omit&&\omit&&\omit&&\omit&\cr}
\hrule}\hfill
\medskip
\centerline{Table 1. Angular eigenvalues for $a/\alpha=0$.}
\medskip
\hfil\vbox{\offinterlineskip
\hrule
\halign{&\vrule#&\strut\quad\hfil$#$\quad\cr
height2pt&\omit&&\omit&&\omit&&\omit&&\omit&&\omit&\cr
&\omega\,a&&(l,m)=(0,0)&&(1,0)&&(1,1)&&(1,-1)&&(2,0)&\cr
height2pt&\omit&&\omit&&\omit&&\omit&&\omit&&\omit&\cr
\noalign{\hrule}
height2pt&\omit&&\omit&&\omit&&\omit&&\omit&&\omit&\cr
&0.0&&0.000&&2.003&&2.021&&2.097&&6.208&\cr
&0.1&&0.007&&2.008&&1.825&&2.310&&6.212&\cr
&0.2&&0.028&&2.020&&1.645&&2.538&&6.228&\cr
&0.3&&0.061&&2.040&&1.482&&2.783&&6.252&\cr
&0.4&&0.109&&2.069&&1.335&&3.044&&6.249&\cr
&1.0&&0.680&&2.473&&0.857&&4.954&&6.694&\cr
height2pt&\omit&&\omit&&\omit&&\omit&&\omit&&\omit&\cr}
\hrule}\hfill
\medskip
\centerline{Table 2. Angular eigenvalues for $a/\alpha=0.1$.}
\medskip
\hfil\vbox{\offinterlineskip
\hrule
\halign{&\vrule#&\strut\quad\hfil$#$\quad\cr
height2pt&\omit&&\omit&&\omit&&\omit&&\omit&&\omit&\cr
&\omega\,a&&(l,m)=(0,0)&&(1,0)&&(1,1)&&(1,-1)&&(2,0)&\cr
height2pt&\omit&&\omit&&\omit&&\omit&&\omit&&\omit&\cr
\noalign{\hrule}
height2pt&\omit&&\omit&&\omit&&\omit&&\omit&&\omit&\cr
&0.0&&0.000&&2.007&&2.079&&2.403&&6.861&\cr
&0.1&&0.008&&2.012&&1.871&&2.628&&6.867&\cr
&0.2&&0.029&&2.025&&1.670&&2.871&&6.881&\cr
&0.3&&0.065&&2.046&&1.508&&3.130&&6.906&\cr
&0.4&&0.116&&2.077&&1.352&&3.407&&6.943&\cr
&1.0&&0.721&&2.700&&1.040&&5.432&&7.375&\cr
height2pt&\omit&&\omit&&\omit&&\omit&&\omit&&\omit&\cr}
\hrule}\hfill
\medskip
\centerline{Table 3. Angular eigenvalues for $a/\alpha=0.2$.}

\end